\documentclass[10pt,epsfig]{article}
\usepackage[cp1251]{inputenc}
\usepackage[dvips]{graphicx}
\usepackage{amsmath}
\usepackage{amssymb}
\usepackage[dvips]{epsfig}   \usepackage[dvips]{changebar}

\def\lsim{\
  \lower-1.2pt\vbox{\hbox{\rlap{$<$}\lower5pt\vbox{\hbox{$\sim$}}}}\ }
\def\gsim{\
  \lower-1.2pt\vbox{\hbox{\rlap{$>$}\lower5pt\vbox{\hbox{$\sim$}}}}\ }

\begin{document}
\title{\textbf{Possible experiment for determination \\ of the role
of microscopic vortex rings \\ in the $\lambda$-transition in
He-II}}
\author{ \textbf{Maksim D. Tomchenko}
\bigskip \\ {\small Bogolyubov Institute for Theoretical Physics,}
  \\ {\small Metrologichna 14-b, 03680 Kiev, Ukraine}}
 \date{\empty}
 \maketitle
 \large
 \sloppy

{\it It is suggested that microscopic vortex rings (MVR) play an
important role in the $\lambda$-transition in helium-II and
substantially determine the value of\/ $T_{\lambda}$. For very
thin films of He-II, with thickness $d$ less than the size of the
smallest MVR, the rings do not fit in and, therefore, do not exist
in such films. Consequently, for superfluid films of He-II, a
peculiarity in the form of a smoothed-out jump should be observed
in the curve $T_{m}(d)$ at the values of thickness approximately
equal to the size of the smallest MVR, $d\approx 6\,\mbox{\AA} \pm
3\,\mbox{\AA}$ ($T_{m}$ is the temperature of the maximum of the
broad peak on the curve of the dependence of the specific heat on
temperature). The absence of a similar peculiarity  will be an
evidence that MVR do not influence the values of $T_{\lambda}$ and
$T_{m}$, and do not play any key role in the $\lambda$-transition.
The currently available experimental data are insufficient for
revealing  the predicted peculiarity.}
\bigskip \\


\section{INTRODUCTION}

The microscopic nature of the $\lambda$-transition in  He-II is
still not quite clear. Most of the authors believe that the
$\lambda$-transition is caused by the destruction of ODLRO and is
accompanied by the exhaustion of the condensate, which probably
has composite nature. The viewpoint according to which the
microscopic vortex rings (MVR) play an important role in the
$\lambda$-transition is also popular enough
\cite{ons}--\cite{lund}.
 The latter idea was proposed about 50 years ago \cite{ons}, but a role of MVR
 in the $\lambda$-transition is not clear until now.  Here, by
microscopic rings we understand vortex rings with radius $R\leq 10\,\mbox{\AA}$
and with quantized circulation $\kappa=\hbar/m$ \cite{pat}.

The superfluid (SF) transition in He-II films on disordered substrates is
characterized by two temperatures, $T_{KT}$ and $T_{m}$ (always
$T_{KT}<T_{m}<T_{\lambda}$).  As the thickness of the SF layer of the film  $d
\rightarrow \infty$, it is observed that $T_{KT}, T_{m} \rightarrow
T_{\lambda}$, where $T_{\lambda}$ is the temperature of the bulk
$\lambda$-transition. The peculiarities at $T\approx T_{KT}$ are caused by the
dissociation of big pairs of the Kosterlitz--Thouless (KT) vortices
\cite{kt}--\cite{nb+}, in particular, a narrow peak is observed at $T\approx
T_{KT}$ in the curve $C(T)$ of the dependence of the specific heat on
temperature \cite{fin,rep}. $T_{m}$ is the temperature of the maximum of the
broad peak (BP) in $C(T)$. This BP may be caused by the dissociation of small
pairs of KT-vortices \cite{nb+}, or it may be a finite-size (FS) rounding of
the $\lambda$-transition \cite{fin,fin+}, or both; it depends on the substrate
and on the value of $d$, see below.

  For thick films with $d \gsim 21\,\mbox{\AA}$, BP correspond to
   FS-rounded $\lambda$-transition \cite{yu}--\cite{gasp04}. At
   $d \lsim 21\,\mbox{\AA}$ (for Nuclepore \cite{yu}), the deviation
   from the scaling law $T_{m}-T_{\lambda}\sim d^{-1/\Theta}$ is
   observed (see Fig.~3 in \cite{yu}), which indicates the appearance of the
   contribution from the KT-vortices (VKT) to BP. With a decrease in
   $d$, the contribution of VKT to BP increases. The contribution of VKT to
   the specific heat, $C_{VKT}$, is proportional  to the
   concentration  of VKT, $N_{VKT}$ (as for MVR, see \cite{fnt}), and $N_{VKT}\sim
   a^{-3}$ \cite{kt}, where $a$ is the core radius of VKT.
   According to \cite{land9,wila}, $a\approx \left (\hbar^{2}d/2mU_{0}n^{(2)}
    \right)^{1/2}$, where $U_0 = \displaystyle\int d\textbf{r}U(r)$, and $U(r)$ includes
 the potential of the substrate. So we have, roughly,
    \begin{equation}
    C_{VKT} \sim N_{VKT} \sim \left (n^{(2)}U_{0}/d \right)^{3/2}.
 \label{0} \end{equation}
  Apparently, the specific heat of the ensemble of VKT depends
 strongly on $d$ and on the substrate potential, and can differ by
 several orders for various substrates and values of $d$. It should be
 expected that $C_{VKT}\sim a^{-3}$ is highest at $d\simeq 1\,\mbox{a.l.}$ (atomic
 layer, $3.6\,\mbox{\AA}$), because $a$ is minimal at
 $d \gsim 1\,\mbox{a.l.}$ \cite{wila}.

Thus, for various substrates and values of $d$, the following versions of
BP in the curve $C(T)$ could be realized:\\
(I)~a single BP caused by dissociation of small pairs of VKT
\cite{nb+}; this is typical for thin films with $d\lsim
1\,\mbox{a.l.}$ (e.g., Millipore- and Anopore-films, $d \approx
1$--$3\,\mbox{\AA}$ \cite{fin+}; for these films BP decreases (in
comparison with the background) with an increase in $d$, at
$d>2\,\mbox{\AA}$,
which signifies the two-dimensional nature of the peak \cite{fin+});\\
(II)~a single BP due to rotons (R) and, perhaps, MVR, wich means
that such a BP is an FS-rounded $\lambda$-peak; this case is
observed for thick films \cite{yu,fss,fred}, e.g., for Nuclepore
at $d\gsim 21\,\mbox{\AA}$
\cite{yu};\\
(III)~a single broad peak (bump) resulting from both VKT and
R+MVR, possible examples of which are Vycor-films at
$d>2\,\mbox{\AA}$ \cite{fin,brew1,brew2}
  and Nuclepore at $d \approx 10$--$20\,\mbox{\AA}$
  \cite{yu,gasp}; in this case BP grows and narrows with an
  increase in $d$, which manifests that BP is an effect of mainly bulk
  quasiparticles, rotons, and MVR; \\
  (IV)~two different BPs at a given $d$, one being caused by VKT and
  the other one being the FS-rounded $\lambda$-peak (this version was
  not observed untill now, but it is possible at $d\sim
  1\,\mbox{a.l.}$, perhaps, also for the substrates of \cite{fin+}).

In our paper, we use the following notation: $T_{m}^{*}$ is the
temperature of BP caused by VKT \cite{nb+,fin+} (with negligible
contribution to BP from rotons and MVR);  $T_{m}$ is the
temperature of the maximum of BP resulting from R+MVR or both VKT
and R+MVR\@. Thus, $T_{m}$ is the temperature of the FS (or
FS+VKT) modified $\lambda$-peak \cite{fin,fin+,brew1,brew2}.

Below, we predict a possible jump, first of all, in the curve
$T_{m}(d)$. A similar peculiarity could also exist in the curve
$T_{m}^{*}(d)$ at $d\simeq d_{0}$ because of the smoothed-out jump
of $\rho_s$ at $d\simeq d_{0}$ (see below), but the BP at
$T_{m}^{*}$ may be undistinguishable at $d\simeq d_{0}$.

 \section{ON THE POSSIBILITY OF A JUMP IN THE CURVE $T_{m}(d)$}
 We suggest that the ensemble of MVR induces the
$\lambda$-transition in the bulk He-II and determines the value of
$T_{\lambda}$  (whatever the mechanism is). This suggestion means that, in the
absence of an ensemble of MVR in He-II, all other quasiparticles would not
cause  the $\lambda$-transition at $T=2.17$~K. To induce the
$\lambda$-transition without MVR, the number of other quasiparticles would have
to grow to a certain critical value. This means that, in the absence of MVR in
the bulk He-II, the value of $T_{\lambda}$ would be higher. The vortex rings
are known to have certain critical size $d_{0}$ \cite{jons}. For thin
superfluid films of He-II, with the thickness $d$ of the superfluid layer less
than the size $d_{0}$ of the smallest MVR, the rings do not fit in and,
therefore, do not exist. Consequently, the value of $T_{m}$ for films with
thickness $d\approx d_{0}$ should abruptly increase in comparison with $T_{m}$
for films with thickness $d$ just larger than $d_{0}$ (in the last case, the
rings still fit in the film). The value of $T_{m}$ is known to decrease with
the decrease of $d$ mainly as a consequence of the finite-size scaling
\cite{yu,fss}, but, as we have shown, a peculiarity should exist in the curve
$T_{m}(d)$ at $d\approx d_{0}$, similar to that shown in Fig.~1.
\begin{figure}
\centering\epsfig{figure=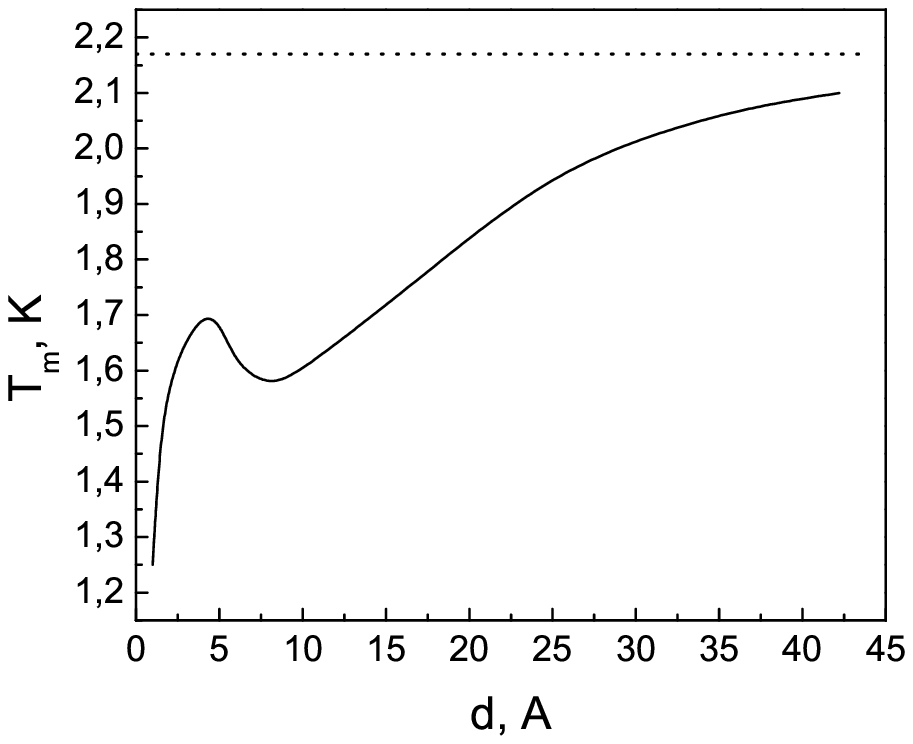,width=0.47\textwidth} \hfill
\centering\epsfig{figure=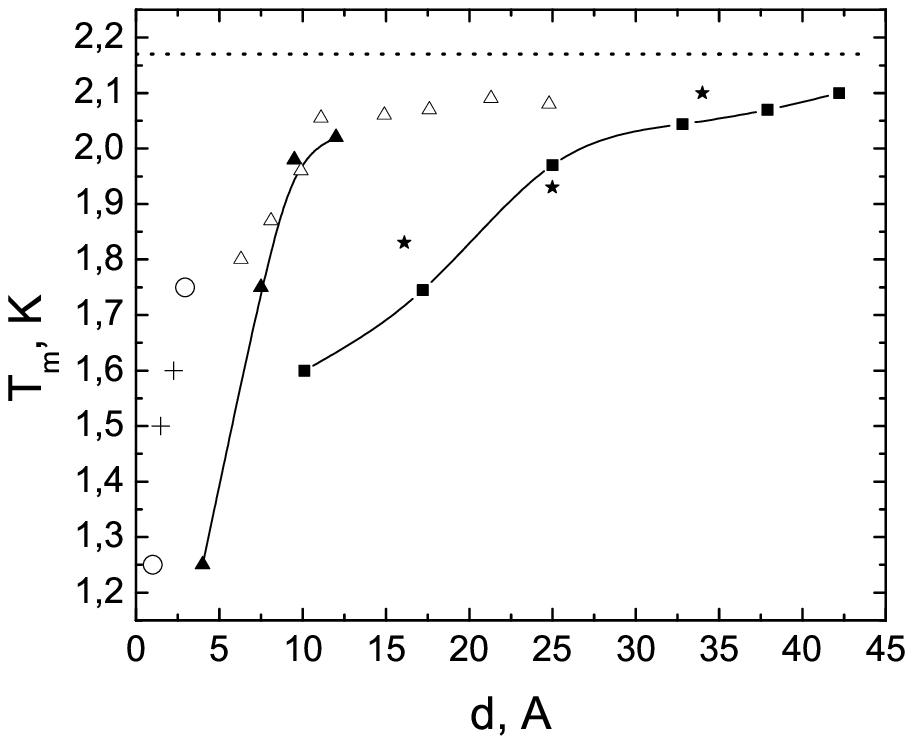,width=0.47\textwidth}
\\
\parbox[t]{0.47\textwidth}{
\caption{The dependence $T_{m}(d)$ for He-II films with the
predicted anomaly at $d\approx d_{0} \simeq 6\,\mbox{\AA}$;  $d$
is  the thickness of the superfluid layer of the film. The values
of $T_{m}$ at $d<2.5\,\mbox{\AA}$ and $d>10\,\mbox{\AA}$
correspond approximately to the crosses and squares in Fig.~2,
respectively, and the dotted line shows $T_{\lambda}$.} } \hfill
\parbox[t]{0.47\textwidth}{
\caption{The experimental dependence $T_{m}(d)$ for He-II films;
$d$ is the thickness of the superfluid layer.  Stars correspond to
the data of Frederikse \cite{fred} with jeweller's rouge
substrate; full triangles correspond to the data of Brewer and
colleagues \cite{brew1,brew2}, Vycor; open triangles correspond to
the data of Brewer \cite{brew2}, $N_{2}$-plated Vycor; squares
correspond to the data of Gasparini and colleagues \cite{gasp},
$2000\,\mbox{\AA}$ Nuclepore; crosses correspond to the data of
Finotello {\it et al.\@} \cite{fin}, Vycor; circles are the points
with $T_{m}=1.25$~K and $T_{m}=1.75$~K from \cite{brew1,brew2},
with $d$ defined more precisely according to \cite{fin}; the
dotted line is $T_{\lambda}$. Continuous lines are drawn by the
spline method.} }
\end{figure}

Experimentally, it is known \cite{ray1,ray2} that vortex rings
with the core radius $a \approx 0.8$--$1.5\,\mbox{\AA}$ may exist
in He-II, and the smallest radius of MVR detected in the
experiment is $R\lsim 5\,\mbox{\AA}$ \cite{ray2}.

Since any vortex should have a core, the size of the smallest
circular or elliptical ring should be roughly two to three core
diameters, $d_{0}\approx 4a$--$6a \approx 3$--$9\,\mbox{\AA}$. For
``lying'' rings, which are parallel to the substrate,
the ``hight" of the ring should be a bit smaller, $d_{0}\approx
3a$ (the core plus the minimal layer of the fluid rotating around
the core). Such dependence of $d_{0}$ on the orientation of the
ring should slightly blur the jump in $T_{m}(d)$.  According to
the approximate model \cite{jons} describing the circular vortex
ring as  a solution of the Gross--Pitaevskii equation,
$d_{0}\approx 4a$, where $a$ is the core radius of the large
rings.

Let us estimate the value of the possible jump of $T_{m}$. We
assume that the $\lambda$-transition in the bulk He-II is
accompanied by complete exhaustion of the one-particle condensate.
According to the calculation \cite{rea}, the fraction of the
one-particle condensate is $n_{0}=0.078$ at $T=0$, and
$n_{0}=0.058$ at $T=T_{\lambda}=2.17$~K, i.e., the condensate does
not vanish completely at $T=T_{\lambda}$, although
$n_{0}(T_{\lambda})\approx 0$ in the experiment. It is suggested
in \cite{rea} that the one-particle condensate in He-II is
exhausted completely [$n_{0}(T_{\lambda})=0$] because of vortex
rings. The rings were not taken into account in the calculation of
\cite{rea}, and the decrease in $n_{0}$ at $T\rightarrow
T_{\lambda}$ was due to rotons: the number of the atoms pulled out
of the condensate was directly proportional to the number of
rotons. The concentration of free rotons is known \cite{land9}:
 \begin{equation} n_{r} =
 0.051*e^{-\Delta/T}\left(\frac{q_{r}}{1.925\ \mbox{\AA}^{-1}}\right)^{2}
 \sqrt{\frac{\mu T_{K}}{0.14m_{4}}}\quad \mbox{\AA}^{-3},
 \label{1} \end{equation}
 where $T_{K}$ is the temperature in Kelvins.
In order that rotons provide $n_{0}=0$, it is necessary that their
number be four times greater than that at $T=2.17$~K; in this
case, we have $n_{0}=0$ instead of $n_{0}=0.058$. For this,
temperature $T\approx 3.12$~K is required according to (\ref{1}).
Thus, if the calculation of $n_{0}$ in \cite{rea} is correct, the
value of $T_{\lambda}$ would be higher, then the observed $2.17K$,
by ${\small \delta} T_{\lambda}=0.95$~K in the absence of the
vortex rings in He-II. This is the upper bound on the value of the
possible jump of $T_{m}$ for He-II films.

  It should be noted that one could expect a small anomaly
 also in the curve $T_{KT}(d)$ at $d\approx d_{0}$. Since the
 rings disappear from the He-II film at $d<d_{0}$, the value of
 $\rho_{s}^{3D}$ should grow at $d\approx d_{0}$ compared to
  $\rho_{s}^{3D}$ at  $d$ just larger than $d_{0}$. So far as
  $T_{KT} \sim d \rho_{s}^{3D}$, a bump-like peculiarity, similar
  to the anomaly in the curve $T_{m}(d)$, should
  exist also in the curve $T_{KT}(d)$ at $d\approx d_{0}$. However, since
  at $d\approx d_{0}$ $T_{KT}$
  is appreciably smaller than $T_{m}$, the number of MVR $N_{vr}$
  and their contribution to $\rho_{s}$ should be several times
  smaller at $T_{KT}$ than at $T_{m}$,
 because  $N_{vr}, \rho_{s}^{vr} \sim exp(-E_{0}/kT)$ ($E_{0}$ is
 the energy of the smallest MVR; the interaction between rings
  may also be important and must be included in $E_{0}$).
  These simple estimates show that the jump in $T_{KT}(d)$ must
  be roughly three to ten times  weaker (or even may be negligible,
 if $E_{0}$ is high enough) than the jump in  $T_{m}(d)$.
 Thus,  one should look for an anomaly, first of all, in the curve
 $T_{m}(d)$.
   The experimental data \cite{agn} do not give clear evidence of the anomaly
 in $T_{KT}(d)$ at $d<7\,\mbox{\AA}$.

Thus, taking into account our estimates and the theory and
experiment for vortex rings, one can see that an ensemble of
microscopic vortex rings, in which the smallest MVR have size
about $d_{0} \simeq 6\,\mbox{\AA}$, should exist in He-II. And, if
the $\lambda$-transition in the bulk He-II is induced by MVR, then
the anomaly should exist in the curve $T_{m}(d)$ at $d \approx
d_{0}\simeq 6\,\mbox{\AA}$. We have drawn this anomaly
approximately in Fig.~1, basing ourselves on the estimates for
$d_{0}$ and $\delta T_{\lambda}$ and taking into account the
smoothing of the jump in $T_{m}$ caused by the finite-size
scaling, by possible heterogeneity of the film thickness and by
the circumstance that part of the rings have size greater than
$d_0$.

It is difficult to make exact calculation of the SF-transition in
thin helium films, taking into account of all kinds of
quasiparticles. But our simple estimates are sufficient for the
prediction of the smoothed-out jump in the curve $T_{m}(d)$ and
for the approximate demonstration of the form and location of the
anomaly.

The experimental data on the dependence $T_{m}(d)$ for thin films
of He-II is shown in Fig.~2. The value of $T_{\lambda}$ for films
is defined as the temperature of the maximum of the broad peak on
the curve of the dependence of the specific heat on temperature.
The crosses in Fig.~2 are obtained using the data of Fig.~1 from
\cite{fin}. According to \cite{fin}, the Brewer's curve
(triangles) should be shifted to the left (circles in Fig.~2); in
this case, the data by Finotello {\it et al.\@} \cite{fin}
(crosses) well agrees with that obtained by Brewer and his
colleagues \cite{brew1,brew2}.  As a whole, as can be seen from
Fig.~2, the data of different works do not fully agree with each
other, and there is a large dispersion of the data points. The
main causes of this disagreement are the imprecise measurement of
the film thickness and the difference in the substrates. Using
these data, we cannot determine the existence of the predicted
peculiarity in the curve $T_{m}(d)$.

More precise measurements of the dependence $T_{m}(d)$ are
necessary for several substrates, for $d$ in the interval between
$1\,\mbox{\AA}$ and $20\,\mbox{\AA}$ with small step $\triangle d
\leq 1\,\mbox{\AA}$. Vycor glass,  Nuclepore and, perhaps, also
Mylar and some other substrates can be used in this case (see
Introduction). These must be substrates on which the He$^4$ films
are superfluid, and KT-effect is observed very well. Ordered
substrates with strong attraction, such as graphite substrates,
are not suitable.
 Precise measurements of $T_{m}^{*}(d)$ and $T_{KT}(d)$
 for $d=1$--$20\,\mbox{\AA}$ could also be interesting.

  A discovery of such an anomaly would be the first experimental
 evidence of the existence of an ensemble of MVR as
  thermal excitations in He-II and could stimulate further theoretical
  and experimental investigations of vortex rings in He-II\@.

The presence of the anomaly will indicate that MVR play an
important role in the bulk $\lambda$-transition and substantially
influence the value of $T_{\lambda}$, and the absence of the
anomaly  will be an evidence that MVR do not influence the value
of $T_{\lambda}$ and do not play any key role in the
$\lambda$-transition in He-II\@. In any case, we would obtain
information on the nature of the $\lambda$-transition in He-II\@.
Therefore, exact measurement of the dependence $T_{m}(d)$ for
He-II films, in our opinion, is of great interest.

\bigskip

The idea of this work is developed in more detail in \cite{fnt}.

\section*{ACKNOWLEDGMENTS}
 The author is
grateful to Yu.\,V.\,Shtanov for valuable discussion.

\renewcommand\refname{REFERENCES}



       \end{document}